**Optomechanical Quantum Control of a Nitrogen Vacancy Center in Diamond**

D. Andrew Golter[1], Thein Oo[1], Mayra Amezcua[1], Kevin A. Stewart[2], and Hailin Wang[1]

[1]Department of Physics, University of Oregon, Eugene, Oregon 97403, USA

[2]School of Electrical Engineering and Computer Science,
Oregon State University, Corvallis, OR 97331, USA

Abstract

We demonstrate optomechanical quantum control of the internal electronic states of a diamond nitrogen vacancy (NV) center in the resolved-sideband regime by coupling the NV to both optical fields and surface acoustic waves via a phonon-assisted optical transition and by taking advantage of the strong excited-state electron-phonon coupling of a NV center. Optomechanically-driven Rabi oscillations as well as quantum interferences between the optomechanical sideband and the direct dipole-optical transitions have been realized. These studies open the door to using resolved-sideband optomechanical coupling for quantum control of both the atom-like internal states and the motional states of a coupled NV-nanomechanical system, leading to the development of a solid-state analog of trapped ions.

Electromagnetic waves have traditionally been the primary experimental tool for controlling a quantum system and for transmitting and distributing quantum information. There has also been strong recent interest in using acoustic or mechanical waves, in particular surface acoustic waves (SAWs), for quantum control and on-chip quantum communication of artificial atoms. Experimental and theoretical efforts have included coherent coupling of SAWs or mechanical vibrations to superconducting qubits[1-3], SAW-based universal quantum transducers[4], strain-mediated coupling between a mechanical resonator and artificial atoms such as nitrogen vacancy (NV) centers in diamond and semiconductor quantum dots (QDs)[5-12], mechanical quantum control of electron spins in diamond[13, 14], phononic QED [4, 15], and phonon-mediated spin squeezing[16].

The most successful exploitation of mechanical vibrations for quantum control, however, combines both optical and mechanical interactions through phonon-assisted optical transitions or sideband transitions, as demonstrated in trapped ions[17-19] and more recently in cavity optomechanics[20, 21]. These optomechanical interactions take place in the resolved-sideband regime, in which the mechanical frequency, $\omega_m$, exceeds the decoherence rate for the relevant optical transitions. For the trapped ion system, the optomechanical processes can control both the internal atomic states and the center-of-mass mechanical motion of an atom. Combining these two aspects of optomechanical quantum control has led to thus far the most successful paradigm for quantum information processing and has also enabled the generation of exotic quantum states such as phonon number states and Schrödinger cat states[17-19]. These remarkable successes have stimulated strong interest in pursuing optomechanical quantum control of artificial atoms such as QDs and NV centers. Ground state cooling and spin entanglement via optomechanical processes in hybrid nanomechanical systems of QDs or NVs have been proposed[22-24]. Resolved sideband emission of a QD coupling to a SAW has also been realized[10].

Here, we report experimental demonstration of optomechanical quantum control of a NV center in diamond through the sideband transitions. We have realized Rabi oscillations of a NV center by coupling the NV simultaneously to both optical and SAW fields in the resolved-sideband regime. Quantum interferences between the optomechanical sideband and the direct dipole-optical (or carrier) transitions have also been observed. These studies represent a major step toward achieving the quantum control of both the internal atom-like states and the motional states of a coupled artificial atom-nanomechanical system. NV centers in diamond have recently emerged as a leading candidate for solid-state spin qubits[25]. With diamond-based micro-electrical-mechanical systems (MEMS), which can be fabricated in a diamond-on-silicon or bulk diamond platform[26-30], the resolved-sideband optomechanical processes demonstrated in our studies can be further used for the quantum control of the mechanical motion[23], unlocking the extensive toolbox of SAWs and MEMS for quantum science and technology.

SAWs, such as Rayleigh waves, propagate along the surface or interface of an elastic material. Mechanical vibrations of SAWs feature both transverse and longitudinal components and extend approximately one wavelength below the surface[31]. These surface waves can be generated electrically in a piezoelectric substrate with the use of inter-digital transducers (IDTs). High frequency SAW devices have previously been fabricated on diamond[32]. For our samples, a 400 nm thick layer of ZnO, which is strongly piezoelectric, was first sputtered onto the diamond surface. IDTs with a designed center frequency near 900 MHz were then patterned on the ZnO surface with electron beam lithography. Details on the fabrication of IDTs and the generation of SAWs by applying a RF signal to an IDT are presented in the supplement[33].

Our experimental studies were carried out at 8 K, with the diamond sample mounted in a cold-finger optical cryostat. A confocal optical microscopy setup enables optical excitation and fluorescence collection of a single NV[34, 35]. An off-resonant green laser beam ($\lambda$=532 nm) is used to initialize the NV into the $m_s$=0 ground state.

For combined mechanical and optical interactions, a NV center situated a few μm below the diamond surface is subjected to an incident laser field and also to a SAW propagating along the diamond surface (see Fig. 1a). An IDT can serve as a transmitter to excite the SAW and also as a detector to characterize the SAW, as shown schematically in Fig. 1b. The $|m_s=0\rangle$ to $|E_y\rangle$ optical transition near $\lambda$=637 nm is used for the optomechanical quantum control (see Fig. 1c)[36-38]. The optical excitation spectrum of this transition for a single NV is shown in Fig. 1d.

The excitation of long-wavelength acoustic phonons in diamond induces a periodic lattice strain. The orbital degrees of freedom of the excited states of a NV couple strongly to this lattice strain, with a deformation potential, $D$, of several eV. The electron-phonon coupling can be characterized by a strain-induced energy shift as well as state mixing of the relevant electronic energy levels[36-38]. For phonon-assisted optical transitions, we consider here the strain-induced energy shift of the NV excited state $|E_y\rangle$, with the electron-phonon interaction Hamiltonian given by[24]:

$$H_{e-phonon} = \hbar g (\hat{b} + \hat{b}^+) |E_y\rangle\langle E_y|, \qquad (1)$$

where $\hat{b}$ is the annihilation operator for the phonon mode, $g = Dk_m\sqrt{\hbar/2m\omega_m}$ is the effective electron-phonon coupling rate, $k_m$ is the wave number of the phonon mode, and $m$ is the effective mass of the mechanical oscillator. With the laser field at the red sideband of the optical transition (see Fig. 1c), the effective interaction Hamiltonian for the first red sideband transition is given by[17],

$$H_R = i\frac{\hbar g \Omega_0}{2\omega_m}(\hat{b}\sigma_+ - \hat{b}^+\sigma_-), \qquad (2)$$

where $\Omega_0$ is the Rabi frequency for the optical field and $\sigma_\pm$ are the raising and lowering operators for the two-level optical transition. The effective Rabi frequency for the sideband transition is thus given by $\Omega = g\sqrt{n}\Omega_0/\omega_m$, where $n$ is the average phonon number. A similar Hamiltonian can also be derived for the first blue sideband transition.

Note that earlier experimental studies on mechanical coupling of NV centers have used exclusively electron-phonon interactions in the ground-state triplet of the NV centers. The ground-state electron-phonon coupling, however, is many orders of magnitude weaker than the excited-state electron-phonon coupling due to the symmetry of the relevant wave functions[37, 38].

We probe the sideband transitions using fluorescence from a NV center driven simultaneously by both optical and acoustic waves. For the photoluminescence excitation (PLE) spectrum shown in Fig. 2a, the NV is initially prepared in the $m_s=0$ state and the fluorescence from the $E_y$ state is measured as a function of the detuning of the incident laser field from the direct dipole-optical transition, with $\omega_m$ fixed at 900 MHz. The blue and red sideband resonances observed in the PLE spectrum correspond to the Stokes and anti-Stokes phonon-assisted optical transitions, respectively, while the carrier resonance at zero detuning corresponds to the direct dipole-optical transition from the $m_s=0$ to $E_y$ states. The spectral separation between the sideband and the carrier resonance equals $\omega_m$, as confirmed by the dependence of the sideband spectral position on the RF driving frequency of the IDT (see Fig. 2b).

At relatively low optical and SAW powers, the peak amplitude of the sideband resonance increases linearly with both the optical and SAW powers (see Figs. 2c and 2d). In comparison with the carrier resonance, the sideband resonances exhibit much weaker saturation and power broadening under the same optical powers, as evidenced by the PLE spectra obtained at three different incident laser powers shown in Fig. 2e. The spectral linewidths of the carrier and sideband resonances are plotted in Fig. 2f as a function of the incident laser power. In the low power limit, the linewidths of the carrier and sideband resonances both approach 175 MHz, which is primarily due to spectral diffusion of the NV center induced by local charge fluctuations from repeated initialization of the NV by the green laser beam. The power broadening of the carrier resonance can be accounted for by a simple two-level model. The broadening also provides a measure of the optical Rabi frequency. From Fig. 2f, we estimate $\Omega_0/(2\pi\sqrt{P_o})=65$ MHz/$\sqrt{\mu W}$, where $P_o$ is the incident laser power[33]. The lack of power broadening for the sideband resonances shown in Fig. 2f indicates that under these experimental conditions, $\Omega$ is still small compared with the NV linewidth.

Excitations from the $m_s$=0 to $E_y$ states can take place through a sideband transition and also through the direct dipole-optical transition (see Fig. 3a), which can lead to quantum interference between these two excitation pathways. To demonstrate this interference, we drive the NV center simultaneously through both the red sideband and the direct dipole-optical transitions. The two optical fields are derived from the same laser, maintaining well-defined relative phase. The optical field near the carrier resonance is generated with an acousto-optic modulator (AOM), which up-shifts the laser field by a frequency, $\omega_{AOM}$. As shown in Fig. 3b, NV florescence as a function of $\omega_{AOM}$ exhibits a sharp resonance when the frequency of the optical field for the direct dipole-optical transition equals the sum of the frequencies of the sideband-detuned optical field and the SAW. The width of the resonance (< 10 Hz) is limited by the instrument resolution. Under this resonant condition, the NV florescence shows a sinusoidal oscillation as we vary the relative phase of the SAW, as shown in Fig. 3c. This oscillation demonstrates the interference between the carrier and sideband transitions and shows that the optomechanical processes are fully coherent with the conventional optical processes. A detailed theoretical analysis of the sharp interference resonance and the sinusoidal oscillation is presented in the supplement[33].

In the resolved sideband limit, we can drive the coherent evolution, in particular, the Rabi oscillations of the two-level NV system using the optomechanical sideband transitions. As illustrated in the pulse sequence shown in Fig. 4a, to realize the Rabi oscillations we tune a continuous optical field onto the red sideband resonance. The acoustic field is turned on for 90 *ns* increments, followed by a 100 *ns* rest time to ensure that the NV center has relaxed back into the ground state, while fluorescence counts are detected and time tagged relative to the beginning of each acoustic pulse. The measurement step is repeated 100 times before a green laser pulse is reapplied.

With the NV initially prepared in the $m_s$=0 state, the florescence from the NV is measured as a function of the acoustic pulse duration. Figure 4b shows the optomechanically driven Rabi oscillations obtained at three different RF driving powers for the IDT. As expected, the effective Rabi frequency, $\Omega$, for the sideband transition derived from the Rabi oscillations is proportional to the square root of the RF driving power (see the inset of Fig. 4b). The oscillations are damped primarily due to spontaneous emission and pure dephasing of the NV center. A detailed theoretical analysis of the Rabi oscillations, including a small background contribution from the carrier transition, is discussed in the supplement[33]. Although higher frequency Rabi oscillations and thus higher fidelity coherent evolution can be achieved, the time resolution of the photon counter (2.8 *ns*) limits the frequency of the Rabi oscillations that can be detected in our current experimental setup.

The Rabi frequency of the sideband transition obtained from these experiments also provides a measurement for the amplitude of the corresponding SAW excitation, given by $A_{SAW} = 2(\omega_m/k_m)(\Omega/\Omega_0)/D$. With $\Omega_0/2\pi$=290 MHz, $\Omega/2\pi$=66 MHz, $\omega_m/k_m$ = 5600 *m/s*, and taking $D/2\pi$=610 THz[24], we estimate $A_{SAW}$=0.7 *pm*, which is in general agreement with the estimated SAW magnitude induced by the IDT, as discussed in the supplement[33]. The relatively small mechanical displacement needed for driving the Rabi oscillations reflects the strong electron-phonon coupling of the NV excited states.

The resolved-sideband optomechanical processes realized in the above experiment can be further exploited for the quantum control of the motional states of a high-Q diamond nanomechanical oscillator that couples to a NV center[23]. For a nanomechanical oscillator with $\omega_m$=900 MHz and mass about 1 *pg*, the excited-state electron-phonon coupling can lead to a single-phonon coupling rate of order 2 MHz. Furthermore, as proposed in an earlier study[24], coherent Raman transitions, such as those used in coherent population trapping of NV centers [34, 35, 39, 40], can be employed to take advantage of both the strong excited-state electron-phonon coupling and the *ms* long ground-state spin decoherence time of NV centers [41].

In summary, we have demonstrated the quantum control of the internal states of a NV center by using optomechanical sideband transitions and by taking advantage of strong excited-state electron-phonon coupling of NV centers. Given the exceptionally low mechanical loss of diamond[26], NV centers provide a highly promising system for combining the quantum control of both the atom-like internal states and the motional states of a coupled NV-nanomechanical system through these optomechanical processes. With extensive MEMS and SAW technologies available for the engineering of nanomechanical systems, NV centers coupling to a nanomechanical oscillator can potentially enable a trapped-ion-like solid-state platform for quantum information processing.

This work is supported by the US National Science Foundation under grants No. 1414462 and No. 1337711.

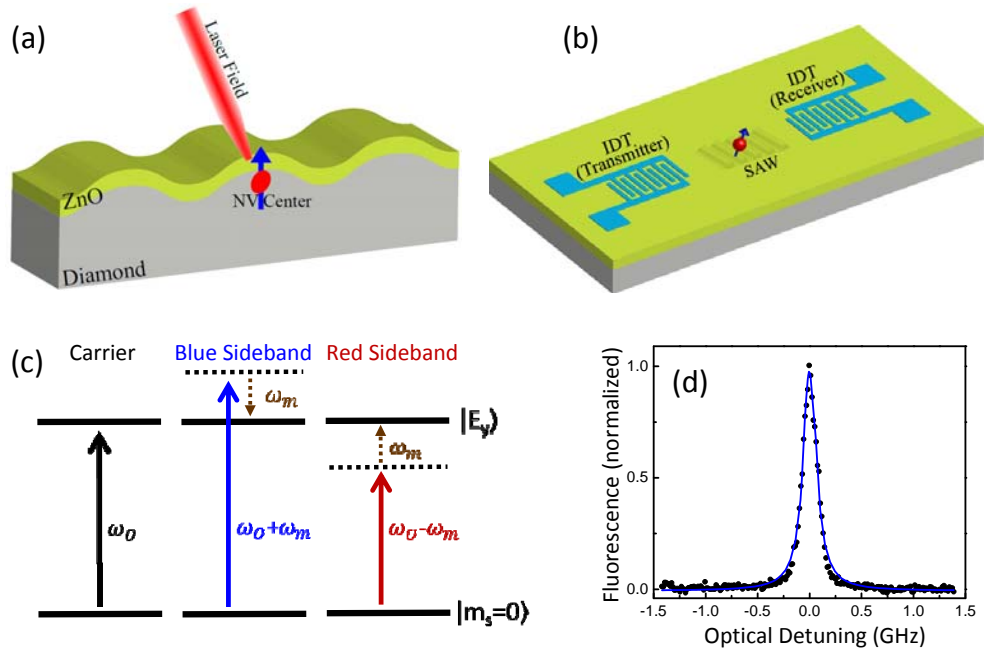

**FIG. 1.** (a) A NV center located near the diamond surface coupling to a laser field and a propagating SAW that extends from the thin ZnO layer to about one acoustic wavelength (≈6μm) beneath the diamond surface. (b) Schematic of the sample illustrating the use of IDTs fabricated on the piezoelectric ZnO layer to generate and detect SAWs. (c) Energy level diagram illustrating the blue and red sideband transitions for the optomechanical interactions. The carrier transition is between the $m_s$=0 ground state and the $E_y$ excited state of the NV. (d) The excitation spectrum of the carrier transition, where NV fluorescence is measured as the laser frequency is tuned across the dipole transition (the acoustic field is off). The blue line is a Lorentzian fit.

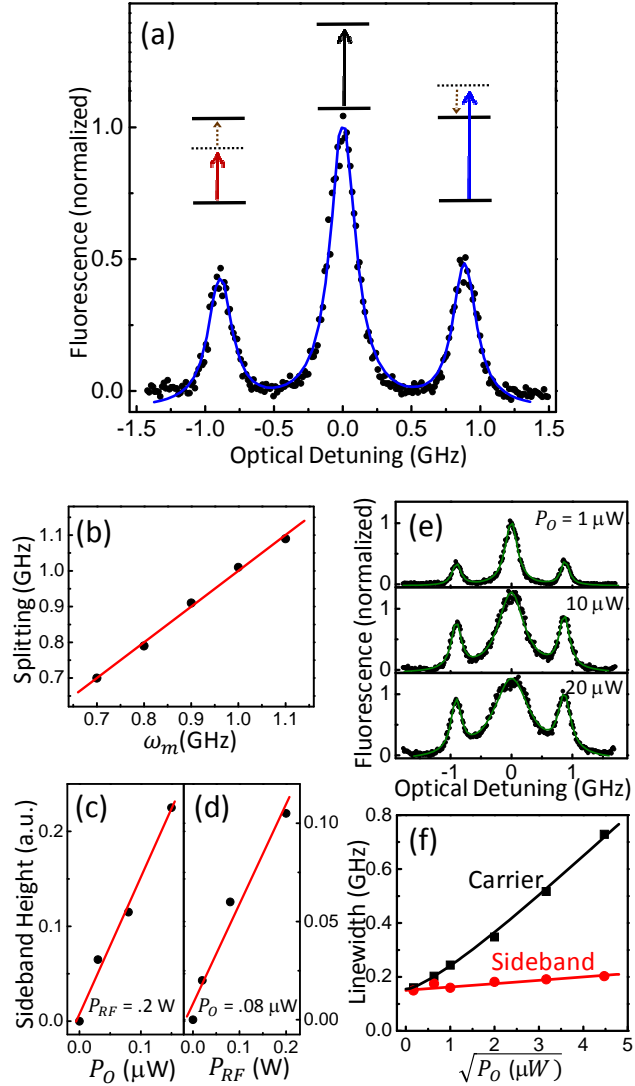

**FIG. 2.** (a) Excitation spectrum with the NV driven by both the optical and acoustic fields and with $\omega_m$=900 MHz. The incident laser power is $P_o = 0.4$ µW and the RF input power is $P_{RF} = 0.2$ W. The blue line is a fit to Lorentzians. (b) Measured frequency splitting between the carrier and the sideband resonances as a function of $\omega_m$. For the red line, the frequency splitting = $\omega_m$. (c) and (d) Amplitudes of the sideband resonances with increasing $P_o$ and $P_{RF}$. Red lines are linear least-squares fits. (e) Excitation spectra of the NV obtained for three different $P_o$ and with $P_{RF} = 0.1$ W. The amplitude is normalized to the peak amplitude of the carrier resonance obtained with $P_o = 0.1$ µW. Green lines are fits to Lorentzians. (f) Linewidths of the carrier (black squares) and red sideband (red circles) resonance as a function of $P_o$, with $P_{RF} = 0.1$ W. Black line is the calculated power broadening. Red line is a liner fit to guide the eye.
12

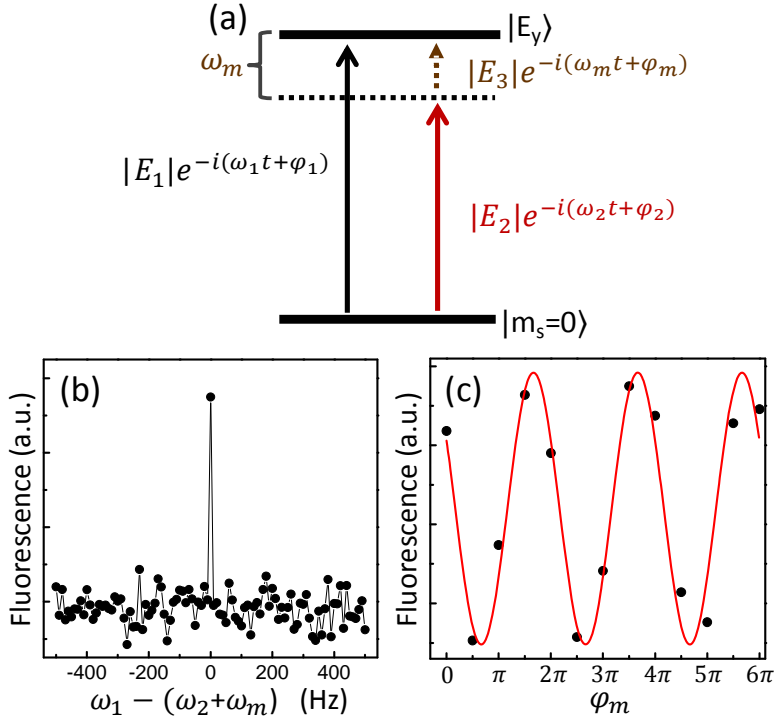

**FIG. 3.** (a) Two pathways, corresponding to the direct dipole transition and the red sideband transition, for the excitation of state $E_y$. The figure also denotes the frequencies and phases of the optical and acoustic waves. (b) NV fluorescence as a function of detuning between $\omega_1$ and $\omega_2 + \omega_m$. (c) NV Fluorescence as a function of $\varphi_m$, with $\omega_1 - (\omega_2 + \omega_m) = 0$, showing the interference between the two pathways. The red line is a sinusoidal oscillation with a period of $2\pi$.

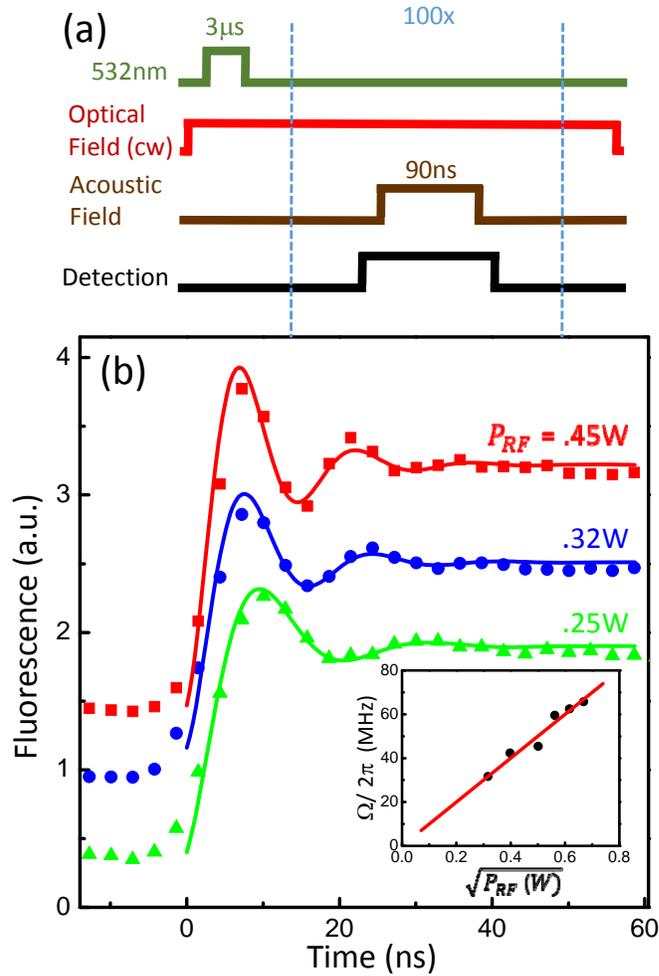

**FIG. 4.** (a) Pulse sequence used for the Rabi oscillation experiment. (b) NV Fluorescence as a function of acoustic pulse duration. Rabi oscillations (offset for clarity) are shown with estimated $\Omega_0/2\pi$ = 290 MHz and for three different RF driving powers for an IDT with $\omega_m$=940 MHz. Solid lines are numerical fits to damped sinusoidal oscillations. Inset: Rabi frequencies obtained as a function of the RF power. Red line shows a linear least-square fit.